%% file: ICC.tex
\documentclass[conference]{IEEEtran}
\input{preamble}

\newif\ifjournal
\journalfalse
\usepackage{algorithm}
\usepackage[numbers,sort&compress]{natbib}
\usepackage[utf8]{inputenc}
\usepackage[noend]{algpseudocode}
\usepackage[english]{babel}
\usepackage{balance}
\def\Ktot{K_{\mathrm{tot}}}
\def\Ka{K_{\mathrm{a}}}

\def\Pr{\mathrm{Pr}}

\usetikzlibrary{arrows,shapes,chains,matrix,positioning,scopes,patterns,calc,decorations.pathmorphing,shadows}
\usepackage[underline=true,rounded corners=false]{pgf-umlsd}
\usepackage{mathtools}
\usepackage{amsmath}
\IEEEoverridecommandlockouts\tikzstyle{cnode}=[circle,draw]
\tikzstyle{cgnode}=[circle,draw]
\tikzstyle{crnode}=[circle,draw]
\tikzstyle{conode}=[rectangle,draw]
\tikzstyle{cpnode}=[circle,draw]
\tikzstyle{rnode}=[rectangle,draw,outer sep=0pt]
\tikzstyle{fnode}=[rectangle,draw,rounded corners,fill=blue!25,align=center]
\tikzstyle{prnode}=[rectangle,rounded corners,fill=blue!50,text width=4.5em,text centered,outer sep=0pt]
\tikzstyle{prnodebig}=[rectangle,rounded corners,fill=blue!50,text width=7em,text centered,outer sep=0pt]
\tikzstyle{prnodesimple}=[rectangle,draw,text width=4.5em,text centered,outer sep=0pt]
\tikzstyle{bigsnake}=[snake=snake,segment amplitude=4mm, segment length=4mm, line after snake=5mm]
\tikzstyle{smallsnake}=[snake=snake,segment amplitude=0.7mm, segment length=4mm, line after snake=3mm]
\tikzstyle{block}=[rectangle,draw,minimum width=2cm,minimum height=1cm,text centered]
\tikzstyle{sblock}=[rectangle,draw,minimum width=1cm,minimum height=1cm,text centered]
\begin{document}
\title{{ Polar Coding and Random Spreading for Unsourced Multiple Access}}
\author{ Asit Kumar Pradhan, Vamsi K. Amalladinne,  Krishna R. Narayanan, and  Jean-Francois Chamberland \\
 Department of Electrical and Computer Engineering, Texas A\&M University
\thanks{
This material is based, partly, upon work supported by the National Science Foundation (NSF) under Grant No.~CCF-1619085.}
}
\maketitle
\begin{abstract}
   This article presents a novel transmission scheme for the unsourced, uncoordinated Gaussian multiple access problem. 
   The proposed scheme leverages notions from single-user coding, random spreading, minimum-mean squared error (MMSE) estimation, and successive interference cancellation.
   Specifically, every message is split into two parts:
   the first fragment serves as the argument to an injective function that determines which spreading sequence should be employed, whereas the second component of the message is encoded using a polar code.
   The latter coded bits are then spread using the sequence determined during the first step.
   The ensuing signal is transmitted through a Gaussian multiple-access channel (GMAC).  
   On the receiver side, active sequences are detected using a correlation-based energy detector, thereby simultaneously recovering individual signature sequences and their generating information bits in the form of preimages of the sequence selection function.
   Using the set of detected active spreading sequences,
   an MMSE estimator is employed to produce log-likelihood ratios (LLRs) for the second part of the messages corresponding to these detected users. 
   The LLRs associated with each detected user are then passed to a list decoder of the polar code, which performs single-user decoding to decode the second portion of the message.  
   This decoding operation proceeds iteratively by subtracting the interference due to the successfully decoded messages from the received signal, and repeating the above steps on the residual signal.
   At this stage, the proposed algorithm outperforms alternate existing low-complexity schemes when the number of active uses is below 225.
\end{abstract}
\section{Introduction}
\label{Intro}
Unsourced multiple access is a novel communication paradigm attuned to machine-type communications (MTC).
Originally proposed by Polyanskiy in~\cite{polyanskiy2017perspective}.
This model forms a significant departure from the traditional information theoretic multiple access channel.
The unsourced random access paradigm seeks to address the distinct nature of the traffic generated by MTC devices.
It captures scenarios where a massive number of devices transmit short payloads in sporadic manner to a central processing unit.
As opposed to traditional multiple access channels, these devices do not produce sustained connections and, hence, the cost involved in transmitting user identities and buffer states cannot be amortized over a long time period.
This leads to a situation where it is beneficial for MTC devices to transmit very short payloads and embed their identity in the data only when they wish to reveal themselves to the central processing unit.
The decoding is then done only up to a permutation of the transmitted messages, impervious to the origin of each message.
This also naturally forces the users to share the same codebook for all their transmissions.
In view of the large number of devices that may wish to transmit at any point in time, we adopt the per-user probability of error (PUPE) performance criterion introduced in~\cite{polyanskiy2017perspective}, as opposed to the stringent global error probability.
Conventional multiple access techniques like ALOHA and treating interference as noise (TIN) are known to be very energy inefficient for the unsourced MAC.
Indeed, there is a significant performance gap between these techniques and the random coding achievability bound, a finite-blocklength (FBL) benchmark derived by Polyanskiy in the absence of complexity constraints~\cite{polyanskiy2017perspective}.

Ever since the introduction of the unsourced MAC challenge, there has been significant effort in designing coding schemes that operate close to the FBL bound while maintaining low computational complexity.
Proposed schemes achieve this by either splitting the payload~\cite{amalladinne2019coded, Giuseppe,amalladinne2019enhanced}, sparsifying collisions~\cite{ordentlich2017low, vem2019user,facenda2019efficient, pradhan2019joint, marshakov2019polar}, or a combination thereof~\cite{calderbank2018chirrup}.
Schemes that rely on splitting the payload take a compressed sensing view of the unsourced MAC problem, and they use a divide-and-conquer approach to limit complexity.
On the other hand, schemes that seek to sparsify collisions employ channel codes that offer good performance in the presence of moderate interference.

In \cite{vem2019user}, the authors present a scheme based on a slotted framework. 
The transmission frame is divided into slots and active devices choose a subset of these slots (based on their message index) to transmit the message. 
Within each slot, devices use an LDPC code designed to perform well in the presence of limited interference. 
Successive interference cancellation (SIC) techniques are applied across slots to cancel the contributions of successfully decoded messages from other slots.
The scheme presented in \cite{marshakov2019polar} resorts to an analogous system architecture.
However, within each slot, a polar code is used to encode device data and detection is performed jointly over all the transmitted codewords in the slot.
Design parameters of the polar code, including frozen bits, are conveyed through a preamble at the beginning of each frame.
This scheme benefits from joint detection at slot level, but becomes increasingly complex to implement when the number of active users is large.
In~\cite{pradhan2019joint}, we describe an adaptation of interleave division multiple access (IDMA) for the unsourced MAC.
Unlike the schemes in \cite{vem2019user,marshakov2019polar}, this design does not rely on a slotted framework.
Each active device picks a sparse pattern and transmits its LDPC coded message over the channel uses specified by the sparse pattern.
Each pattern is picked based on a portion of the corresponding payload, and they are conveyed to the decoder using a compressed sensing scheme.
The decoder employs soft message passing rules to jointly recover the payloads.
The latter two schemes discussed above represent the state-of-the-art for the unsourced MAC paradigm in terms of gap from the FBL bound.

In this article, contrary to the aforementioned approaches, we use random spreading as a means to mitigate multi-user interference.
The device payload, before spreading, is encoded using a polar code designed to perform well in the absence of interference.
This approach enables the application of single user decoding as opposed to schemes which use joint detection~\cite{marshakov2019polar} and, as such, it simplifies the recovery process remarkably.
Also, we implement an energy detector to identify the spreading sequences employed by the active users.
This allows us to do away with the compressed sensor completely and, hence, no channel resources are dedicated exclusively for spreading sequence detection.
The decoding algorithm performs multiple iterations on the received signal by canceling the contribution of successfully decoded signals in the spirit of successive interference cancellation.
Through a numerical study, we demonstrate that the proposed approach outperforms the schemes in \cite{pradhan2019joint,marshakov2019polar} in pertinent regimes.

We use the following notation throughout this article.
The sets $\mathbb{R}$, $\mathbb{Z}$ denote the real numbers and the integers, respectively. 
We write $[a:b]$ as a shorthand notation for $\{c \in \mathbb{Z}:a \le c \le b\}$.
The operator $\otimes$ symbolizes a tensor product.
The conventional $\lfloor x \rfloor$ and $\lceil x \rceil$ are the floor and ceil functions of real number $x$, respectively.
The $\ell_2$ norm of vector $\xv$ is expressed as $\| \xv \|$.
We refer to the $k$th element of vector $\xv$ as $\xv(k)$.
For two vectors $\xv, \yv \in \mathbb{R}^n$, the operator $\langle\xv, \yv\rangle$ denotes the standard inner product.
Finally, we write $\mathbf{V}_{m,:}$ and $\mathbf{V}_{:,l}$ to represent the $m$th row and $l$th column of matrix $\mathbf{V}$, respectively.

\section{System Model}

In the considered random access paradigm, $\Ka$ users out of $\Ktot$ possible users are active at any given time, and each active user wishes to transmit $B$ bits of information to an access point in $n$ uses of the channel.
Let $s_i$ be an indicator random variable which is $1$ if the user $i$ is active, and $0$ otherwise.
The received signal, $\yv$, at the access point is given by
\begin{equation}
\yv = \sum_{i=1}^{\Ktot} s_i \xv_{i}(\underbar{w}_{i}) + \zv,
\end{equation}
where $\underbar{w}_{i} \in \{0,1\}^B$ corresponds to the $B$-bit message user $i$ intends to communicate to the access point, $\xv_{i} \in \mathbb{R}^n$ denotes the signal transmitted by the $i$th user, $\zv \sim \mathcal{N}(0,\mathbf{I}_n)$ is the additive white Gaussian noise (AWGN), and $\sum_i^{\Ktot}s_i=\Ka$.
It is assumed that when $s_i=1$, user~$i$ chooses its message $W_i$ uniformly from  the set $[M] \triangleq [1:M]~ (M=2^B)$. It is further assumed that  messages chosen by the active users are independent. User transmissions need to satisfy a power constraint $\| \xv_{i} \|^2 \leq nP$, for $i \in [1:\Ktot]$.
The energy-per-bit of the system is defined as $\frac{E_b}{N_0} \coloneqq \frac{nP}{2B}$. The decoder produces a list, $\mathcal{L}(\yv)$, of messages with size at most $\Ka$.
The per-user probability of error of the system is given by
\begin{equation}\label{eqn:proboferrordefinition}
  P_e = \max_{\sum s_i = \Ka} \frac{1}{\Ka} \sum_{i=1}^{\Ktot} s_i \Pr\left( w_i \notin \mathcal{L}(\yv) \right).
\end{equation}
For fixed values of $n$, $B$, $\Ka$, $\epsilon$, the objective is to design a low-complexity coding scheme which achieves $P_e \leq \epsilon$, where $\epsilon$ is a target error probability at low ${E_b}/{N_0}$.

\section{Description of Proposed Scheme}
\label{Sec:ProposedScheme}

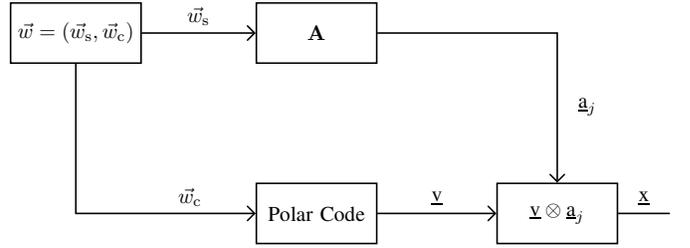
\begin{figure}
    \centering
    \input{encoder.tex}
    \caption{This notional diagram offers an illustration of the encoding process for the proposed scheme.}
    \label{fig:encoder}
\end{figure}

We begin this section with a description of the encoding process.
We then turn to the more intricate decoding algorithm.
A notational diagram for the proposed scheme can be found in Fig.~\ref{fig:encoder}.

\subsection{Encoder}
For a specific message, the $B$ bits to be transmitted are partitioned into two parts of size $B_\mathrm{s}$ and $B_{c}=B-B_\mathrm{s}$ bits, respectively.
For notational convenience, we define $M_s=2^{B_s}$ and $M_c=2^{B_c}$.
We denote the first and second part of the message corresponding to user~$i$ by $\underbar{w}_{\mathrm{s}}$ and $\underbar{w}_{\mathrm{c}}$; that is, $\underbar{w}=(\underbar{w}_{\mathrm{s}}, \underbar{w}_{\mathrm{c}})$. 
The signature sequences utilized by the active users are determined as follows.
Let $\mathbf{A} =[\av_1,\av_2,\ldots,\av_{M_s}]\in \mathbb{R}^{n_s \times M_s}$ denote the codebook of possible sequences.
The elements of $\mathbf{A}$ are created by drawing independent, zero mean Gaussian random variables with unit variance, and rescaling the vectors to satisfy the power constraint.
Every active user is assigned one column of matrix $\mathbf{A}$ as its signature sequence based on its preamble $\underbar{w}_s$ bits.
In other words, each active user employs function $f:\{0,1\}^{B_s}\rightarrow\{\av_j:j \in [1:M_s]\}$ (common to all users) to map preamble bits to columns of the sequence codebook $\mathbf{A}$.
We emphasize that $f$ is bijective almost surely.
For ease of exposition, we denote the spreading sequence picked by active user~$i$ as $\av_{j_i}$
and the length of the polar code by $n_\mathrm{c} = \lfloor n/{n_{\mathrm{s}}} \rfloor$.
The second part of the message, $\underbar{w}_c$, serves as the argument to the polar encoding.
Note that, before polar encoding, we generate and append $r$ cyclic redundancy check (CRC) bits to $\underbar{w}_c$.
The CRC bits are eventually leveraged as statistical evidence of successful polar decoding. 
The resulting sequence of bits $\underbar w_\mathrm{c}$ is first encoded into an $n_\mathrm{c}$-bit codeword $\underbar v'$ of an $(n_\mathrm{c},B_{\mathrm{c}}+r)$ polar code.
The polar codeword $\underbar v'$ is then modulated using binary phase shift keying (BPSK) to generate vector $\underbar v=\{v(1),v(2),\cdots,v(n_\mathrm{c})\}$.
Finally, each symbol of the modulated codeword is multiplied by a spreading sequence to produce the transmitted signal.
Altogether, the signal transmitted by active user~$i$ takes the form
\begin{equation} \label{eq:3}
    \xv_i= \underbar v_i \otimes \av_{j_i},
\end{equation}
where $v_i$ is the modulated polar codeword and $\otimes$ denotes the tensor product operation.
We note that signals generated this way satisfy the power constraint $\| \xv_i \|^2 \leq P$ for any information message.

In summary, the encoding operation is characterized by two broad components.
The first aspect includes encoding with a polar code followed by modulation.
The second component is random spreading, which helps the receiver address some of the challenges posed by the unsourced MAC, while also limiting decoding complexity, as we will see shortly.

\subsection{Decoder}

The decoding process features an iterative structure, with two distinct stages.
During the initial stage, an energy detector is used to identify the set of spreading sequences employed by the active users. 
During the subsequent step, a minimum mean square error (MMSE) estimator produces soft estimates of the symbols corresponding to the detected sequences.
These estimated symbols are then passed to the list decoder of the polar code.
If list decoding is successful, the signals corresponding to the recovered codewords are removed from the received signal in the spirit of SIC.
The residual signal is then redirected to the energy detector. 
This iterative decoding process continues until all the transmitted messages are recovered, or the number of decoded messages does not improve between two consecutive iterations.

\subsubsection{Energy Detector}

As its name suggests, the energy detector seeks to identify active sequences based on a statistics that incorporates energy.
One difficulty in implementing this algorithm stems from the fact that $\xv_i= \underbar{v}_i \otimes \av_{j_i}$ and, at this stage in the decoding process, $\underbar{v}_i$ is unknown.
One naive approach would be to correlate $\yv$ with all the possible vectors of this form; however, this approach is computationally impractical.
A viable alternative is to section $\underbar{v}_i$ into groups
\begin{equation*}
    \underbar{v}_i = \underbar{v}_i(1:g) \underbar{v}_i(g+1:2g) \cdots \underbar{v}_i(n_c-g+1:n_c)
\end{equation*}
and build a decision statistics based on correlating each group with all possible columns in $\mathbf{A}$ and all admissible subvectors of $\underbar{v}_i$.
This yields a statistics of the form
\begin{equation*}
    \sum_{k} \left| \left\langle \yv((k-1)g n_s + 1: k g n_s), \underbar{b} \otimes \av_{j} \right\rangle \right|
\end{equation*}
where $\underbar{b} \in \{ -1, 1 \}^g$ and $\av_{j} \in \mathbf{A}$.
There is a natural tradeoff between the size of each group and the complexity of running this energy detector.
Larger groups permit noise averaging, yet the number of possible $\underbar{b}$ increases.
Correspondingly, smaller groups are easy to manage in terms of complexity, but are more prone to errors due to noise.

Based on this statistics, every column in $\mathbf{A}$ is sorted in descending order.
The energy detector then outputs the first $K$ sequences of the sorted list, where $K=\Ka+K_{\delta}$ for some fixed small non-negative integer $K_{\delta}$. 
The reader may note that the first $B_s$ bits used to pick a spreading sequence by the active users are implicitly decoded by the energy detector.

\subsubsection{Demodulator and Channel Decoder}
\begin{figure*}
    \centering
    \input{decoder.tex}
    \caption{The iterative decoding process involves several stages including energy detection (ED), MMSE estimation, single-user decoding of the polar code, and successive interference cancellation (SIC).}
    \label{fig:decoderl}
\end{figure*}
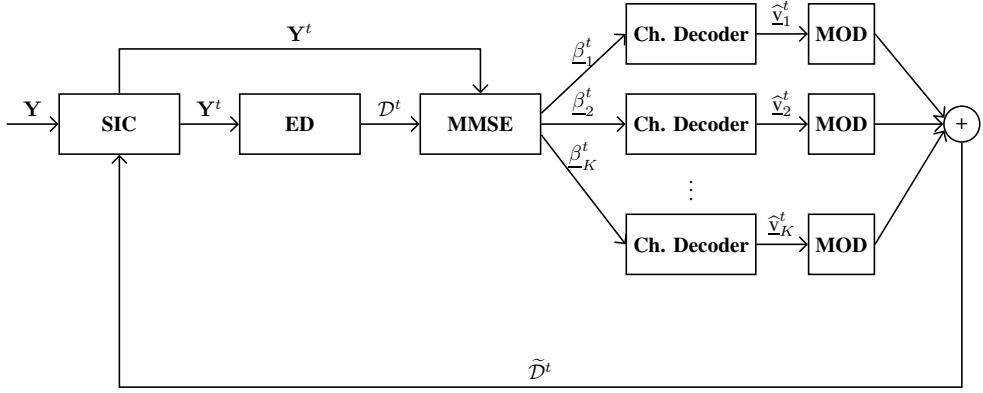
In this section, we describe the demodulation and channel decoding operations.
The exposition is presented for the first iteration of decoding.
Still, the reader may observe that this explanation is valid for all iterations when the received signal $\yv$ is replaced by the residual signal pertaining to that iteration.
We denote the set  of sequence indices returned by the energy detector by $\mathcal{D}$. 
The modulated polar codewords corresponding to all the active users can be stacked in the form of a matrix, given by
\begin{equation*}
    \mathbf{V} \coloneqq \begin{bmatrix}\underbar v_1\\
    \underbar v_2\\
    \vdots\\
    \underbar v_{K}
    \end{bmatrix}.
\end{equation*}
Also, we denote the received signal corresponding to the transmissions of the $l$th polar coded symbol by
\begin{equation*}
    \yv_l= \left[ y((l-1)n_{\mathrm{s}}+1) \; y((l-1)n_{\mathrm{s}}+2) \; \cdots \; y(ln_{\mathrm{s}})\right]^{\mathsf{T}}.
    \end{equation*}
We define a matrix $\mathbf{Y}$, which is a reshaped version of the received vector $\yv$, by
$$\mathbf{Y}= \Big[ \yv_1 \; \yv_2 \; \cdots \; \yv_{n_{\mathrm{c}}}\Big].$$ 
It is straightforward to verify that matrix $\mathbf{Y}$ can be expressed as
\begin{equation}
    \label{eq:5}
    \mathbf{Y}=\mathbf{A}_{\mathcal{D}}\mathbf{V}+\mathbf{Z},
\end{equation}
where $\mathbf{A}_{\mathcal{D}}$ is matrix $\mathbf{A}$ restricted to the columns indexed by elements in the set $\mathcal{D}$.
At this point, we make a simplifying approximation and take the entries of matrix $\mathbf{Z}$ to be independent zero mean Gaussian random variable with unit variance (see Remark \ref{Remark1}).
The covariance matrix of vector $\yv_{l}$ then becomes
\begin{equation}
    \label{eq:6}
    \mathbf{R}=(\mathbf{A}_{\mathcal{D}}\mathbf{A}_{\mathcal{D}}^\mathsf{T}+{I}_{n_\mathrm{s}}).
\end{equation}
We evaluate the MMSE filtering of $\mathbf{Y}$ to get a linear estimate of $\mathbf{V}$ given by
\begin{equation}
    \label{eq:7}
    \hat{\mathbf{V}}=\begin{bmatrix}
    \widehat{\underbar v}_1\\
    \widehat{\underbar v}_2\\
    \vdots\\
    \widehat{\underbar v}_K
    \end{bmatrix}=\mathbf{F}\mathbf{Y},
\end{equation}
where $$\mathbf{F}=\mathbf{A}^{\mathsf{T}}_{\mathcal{D}}\mathbf{R}^{-1}.$$ 
The mean square error (MSE) of this estimator is well approximated by
\begin{equation}
    \label{eq:8}
   \mathbf{\Sigma}=\mathbf{A}^\mathsf{T}_\mathcal{D}\mathbf{R}^{-1}\mathbf{A}_\mathcal{D},
\end{equation}
where $\mathbf{\Sigma}=\operatorname{diag} \left( \sigma^2_{\mathrm{mse}}(1), \sigma^2_{\mathrm{mse}}(2), \ldots, \sigma^2_{\mathrm{mse}}(K) \right)$.
In this context, $\sigma^2_{\mathrm{mse}}(i)$ captures the MSE experienced by the message of user~$i$.
Assuming the MSE is Gaussian, the MMSE estimate of symbol $\underbar v_{i}(l)$ can be seen as the output of an equivalent AWGN channel with noise variance $\sigma^2_{\mathrm{mse}}(i)$ and is given by
\begin{equation}
    \label{eq:9}
    \widehat{ \underbar v}_{i}(l)=\underbar v_{i}(l)+\zeta,
\end{equation}
where $\zeta \sim \mathcal{N}(0,\sigma^2_{\mathrm{mse}}(i))$. 
Then, the LLR of symbol $\underbar v_{i}(l)$ at the output of the equivalent AWGN channel takes the form
\begin{equation*}
\beta_{i}(l)=2\widehat{\underbar v}_{i}(l)/\sigma^2_{\mathrm{mse}}(i),
\end{equation*}
where $\space  1 \leq l \leq n_{\mathrm{c}}$ and  $1 \leq i \leq K.$
The LLR vectors corresponding to each codeword are passed to the single-user list decoder of the polar code. 
The list decoding of the polar code is declared successful if the codeword returned by the decoder satisfies the CRC checks.
    We denote by $\widetilde{\mathcal{D}}$, the collection of indices such that $\underbar{v}_i$ for each $i \in \widetilde{\mathcal{D}}$ is decoded successfully.

The SIC removes the contributions from all the successfully decoded codewords $\mathbf{V}_{\widetilde{\mathcal{D}}}$ from the received signal to compute the residual
\begin{equation*}
\mathbf{Y} -\mathbf{A}_{\widetilde{\mathcal{D}}}\mathbf{V}_{\widetilde{\mathcal{D}}} .
\end{equation*}
This residual is passed back to the energy detector for the second iteration.
This process continues until all the transmitted messages are recovered successfully or there is no improvement between two consecutive rounds of iterations.
We encapsulate the overall decoding process in Algorithm~\ref{alg:decoding}, in which $t$ marks the decoding iteration. 
\begin{remark}
\label{Remark1}
The energy detector gives a subset of spreading sequences picked by the active users in each iteration. 
Hence, \eqref{eq:5} is only an approximate representation of $\mathbf{Y}$, since the undetected sequences are not accounted for.
This effect is more pronounced in the first few rounds of iterations.
However, empirically, we observed that this approximation does not affect the error performance adversely and this is noted in Fig.~\ref{fig:resultsICC}.
\end{remark}

\begin{algorithm}
\caption{Decoding algorithm}\label{alg:decoding}
\begin{algorithmic}[1]
\State Initialize $t=0,\widetilde{\mathcal{D}}^{0}=\varnothing$.
\State Update $t=t+1$.
\State  Initialize $\widetilde{\mathcal{D}}^{t}=\varnothing$.
\State Energy detector returns $\mathcal{D}^t$ such that for each $j \in \mathcal{D}$, $e_j \geq e_{j'},~\forall~j' \in [1:2^{B_{\mathrm{s}}}] \setminus \mathcal{D}^t$.
\State Compute $\widehat{\mathbf{V}}^t=\mathbf{A}_{\mathcal{D}^t}^\mathsf{T}(\mathbf{A}_{\mathcal{D}^t}\mathbf{A}_{\mathcal{D}^t}^\mathsf{T}+{I}_{n_\mathrm{s}})^{-1}\mathbf{Y}^t$.
\State Compute empirical MSE $(\sigma_{\text{mse}}^t(i))^2,~\forall~ i \in \mathcal{D}.$
\State Compute $\beta_i^t(l)=2v_i^t(l)/(\sigma_{\text{mse}}^t(i))^2,~\forall~l \in [1:n_{\mathrm{c}}].$
\State Pass $\underline{\beta}_i^t$ to the list decoder.
\State Update $\widetilde{\mathcal{D}}^t$ as the set of  indices corresponding to successfully decoded codewords.
\State Update $\mathbf{Y}^{t+1}=\mathbf{Y}^t -\mathbf{A}_{\widetilde{\mathcal{D}}^t}\mathbf{V}_{\widetilde{\mathcal{D}}^t}.$
\State Repeat steps 2--9 until $\widetilde{\mathcal{D}}^t = \varnothing$ or $|\bigcup\limits_{t} \widetilde{\mathcal{D}}^t|=\Ka.$
\end{algorithmic}
\end{algorithm}

\section{Simulation Results}

To facilitate a fair comparison between other existing schemes tailored to the unsourced MAC and the proposed approach, we use the following parameters for numerical simulations.
The number of active users $\Ka \in [25:300]$ and each user transmits a payload consisting of $B=100$ bits.
These bits are encoded into $n=30000$ channel uses and transmitted into the channel.
The target per-user error probability is $P_e=0.05$.
The rate of the polar code is then given by $R=\frac{B-B_s}{\lfloor{n /n_s}\rfloor}$.
The spreading sequences are picked from a common codebook $\mathbf{A} \in \mathbb{R}^{n_s \times M_s}$.
Since active users pick columns independently from $\mathbf{A}$, the event in which two or more users pick the same column, which we refer to as a collision event, occurs with a non-zero probability and this probability can be controlled by changing $B_{\mathrm{s}}$. 
Below, we explain briefly how such collisions are handled in the proposed scheme.
Let $\underbar m \in \mathbb{Z}^{K}$ be a vector whose $i$th element $\underbar m(i)$ denotes the number of active users that pick spreading sequence ~$\av_i \in \mathcal{D}$.
In the absence of noise, $\widehat{\underbar v}_i$, the MMSE estimate of $\underbar v_i$, can be seen as the output of a $\underbar m(i)$-real adder MAC.  
We choose the value of $B_{\mathrm{s}}$ to ensure that the probability of the event $\underbar m(i) > 2$ is negligible for all $\av_i \in \mathcal{D}$. 
Assuming $\sigma^2_{\text{mse}}(i)=0$, in the case where $\underbar m(i)=2$, the MMSE estimate $\widehat{\underbar v}$ can be seen as the output of an erasure channel with erasure probability $0.5$.
Since the rate of the polar codes used is less than $1/2$, the single-user list decoder is able to successfully decode even when $\underbar m(i)=2$, at the later stages of the SIC process when most of the interference from other users has been cancelled.

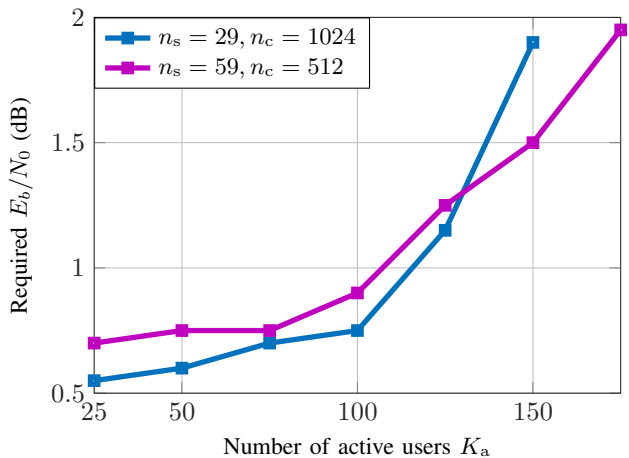
\begin{figure}
    \centering
    \input{comparison.tex}
    \caption{The plot compares the minimum $E_b/N_0$ required by the proposed scheme as a function of $\Ka$ for different values of $n_{\mathrm{s}}$, and $n_{\mathrm{s}}$. The optimal values for $n_{\mathrm{c}}$ and $n_{\mathrm{s}}$ change as functions of $\Ka$.}
    \label{fig:comparison}
\end{figure}

The allocation of channel uses to the polar code and spreading sequences has a significant effect on the performance of the system. 
Also, the optimal values of $n_{\mathrm{c}}$ and $n_{\mathrm{s}}$ change with $\Ka$. 
For example, for $\Ka=150, P_e=0.05$ an $E_b/N_0$  of $1.45$~dB is required when $n_{\mathrm{c}}=512, n_{\mathrm{s}}=59$, whereas an $E_b/N_0$  of $1.9$~dB is required when $n_{\mathrm{c}}=1024, n_{\mathrm{s}}=29$.
For a fixed value of $\Ka$, the values of $n_{\mathrm{c}}$ and $n_{\mathrm{s}}$ are optimized empirically to minimize the $E_b/N_0$ required to achieve a target probability of error. 
The minimum $E_b/N_0$ required to achieve a target probability of error for different values of $n_{\mathrm{c}}$ and $n_{\mathrm{s}}$ as function of $\Ka$ is plotted in Fig. \ref{fig:comparison}.  

\begin{table}
\begin{center}
\begin{tabular}{ |c|c|c|c|c|c|c|c| } 
 \hline
 {$\Ka$} &$B_{\mathrm{s}}$ &$n_{\mathrm{c}}$ & $B_{\mathrm{c}}$ & $n_{\mathrm{s}}$ & {List size} & {$r$}\\
 \hline
 $25-125$ & $9$ & $91$ & $1024$ & $29$ & $1024$ & $16$ \\
 \hline 
 $150$ & $10$ & $90$ & $512$ & $59$ & $128$ & $12$ \\
 \hline
 $175-250$ & $12$ & $88$ & $256$ & $117$ & $128$ & $10$ \\
 \hline
\end{tabular}
\end{center}
\caption{This is a summary of the encoding parameters used in this article for as functions of the number of active users.}
\label{tab:code_parameters}
\end{table}

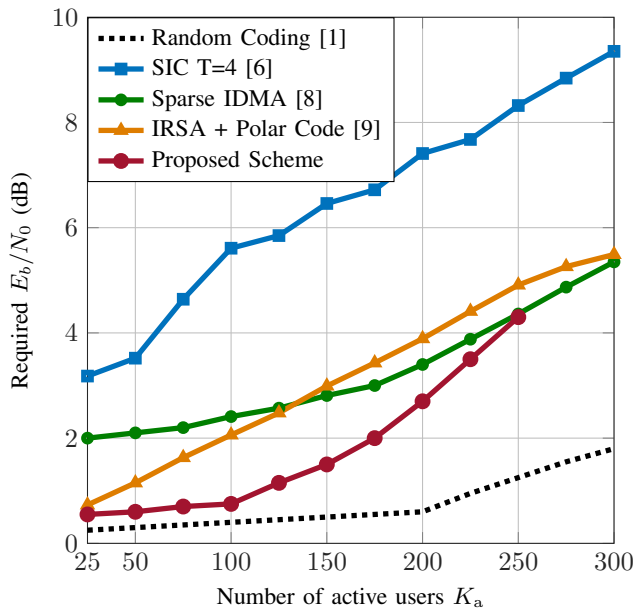
\begin{figure}
\centering
  \input{results_ICC.tex}
  \caption{The figure compares the performance of the proposed scheme with existing schemes. The proposed scheme outperforms the state-of-the-art when $\Ka \le 225$.}
  \label{fig:resultsICC}
\end{figure}

Fig.~\ref{fig:resultsICC} demonstrates the performance comparison between the proposed schemes and previously published methods in the literature.
The parameters used, for these simulations, for different values of $\Ka$  are given in Table \ref{tab:code_parameters}.
The obtained simulation results show that the proposed scheme outperforms existing approaches when $\Ka \leq 225$. 
For example when $\Ka=100$, the proposed scheme outperforms the state-of-the-art~\cite{marshakov2019polar} by $1.31$~dB. 
For $\Ka \leq 100$, the simulated performance is only $0.35$~dB away from the FBL achievability bound developed in \cite{polyanskiy2017perspective}.
We remark that performance can be further improved by carefully optimizing the parameters when $\Ka>200$.

\section{Conclusion}
We presented a coding scheme based on random spreading, single user decoding and interference cancellation for the unsourced multiple access channel.
The proposed scheme uses spreading sequences with good correlation properties to mitigate multi-user interference.
Polar code with list decoding is used as the choice channel code for single user coding.
Simulation results demonstrate that the proposed scheme outperforms existing schemes in some regimes and represents the state-of-the-art for $\Ka \le 225$.
The spreading sequences employed in this work are random Gaussian sequences.
It would be interesting to verify if well designed spreading sequences would be beneficial over random Gaussian sequences in the regime of interest.
Also, the error performance of this scheme is heavily dependent on the lengths of spreading sequence and channel code used.
A key question in this context would be the optimization of these lengths that would help us exploit this trade-off.
Introducing sparsity to user transmissions in the spirit of \cite{pradhan2019joint} might help the decoding algorithm in cases where the system suffers from heavy interference.

\balance
\bibliographystyle{IEEEbib}
\bibliography{IEEEabrv,MACcollision}
\end{document}

%% file: preamble.tex
\usepackage{amsmath,amssymb,amsthm}
\usepackage{mathtools}
\usepackage{enumerate}
\usepackage{stfloats}
\usepackage{comment}
\usepackage{subfig}

\usepackage[dvipsnames]{xcolor}
\definecolor{myPink}{RGB}{255,105,183}

\usepackage[T1]{fontenc}
\usepackage{graphics} 
\usepackage{epsfig} 
\usepackage[mathscr]{euscript}
\usepackage{algorithm}
\usepackage[noend]{algpseudocode}
\makeatletter
\def\BState{\State\hskip-\ALG@thistlm}
\makeatother

\usepackage{hyperref}

\usepackage{tikz}
\usetikzlibrary{arrows,shapes,chains,matrix,positioning,scopes,patterns,calc,
decorations.markings,
decorations.pathmorphing,
}

\usepackage{pgfplots}
\pgfplotsset{compat=1.3}
\usepgflibrary{shapes}

\newtheorem{theorem}{Theorem}

\newtheorem{remark}[theorem]{Remark}

\renewcommand{\epsilon}{\varepsilon}

\newcommand{\RNum}[1]{\uppercase\expandafter{\romannumeral #1\relax}}

\newcommand{\av}{\vec{{a}}}

\newcommand{\wv}{\vec{{w}}}

\newcommand{\xv}{\vec{{x}}}

\newcommand{\yv}{\vec{{y}}}

\newcommand{\zv}{\vec{{z}}}

\def\Ktot{K_{\mathrm{tot}}}
\def\Ka{K_{\mathrm{a}}}

\DeclareMathAlphabet{\mcl}{OMS}{cmsy}{m}{n}

\newlength\tikzwidth
\newlength\tikzheight



%% file: encoder.tex
\begin{tikzpicture}[every node/.style={scale=0.8}]
\begin{scope}[node distance=2cm,>=angle 90,semithick]
\node[block] (message) {$\wv=(\wv_{\mathrm{s}},\wv_{\mathrm{c}})$};
\node[block] (A)[right of=message,xshift=2cm,yshift=0cm]{$\mathbf{A}$}; 
\node[block] (chcode)[below of=A,xshift=0cm,yshift=-1cm]{{Polar Code}}; 
\node[block] (spreading)[right of=chcode,xshift=2cm]{$\underbar v \otimes \underbar a_j$};
\node (output)[right of=spreading,xshift=0cm]{};
\draw[->] (message) -- node[above]{$\wv_{\mathrm{s}}$} (A);
\draw[->] (message.270)  |- node[above,xshift=1.9cm]{$\wv_{\mathrm{c}}$} (chcode);
\draw[->] (A.0)  -| node[above,xshift=0.5cm,yshift=-1.5cm]{$\underbar a_j$} (spreading);
\draw[->] (chcode) -- node[above]{$\underbar v$} (spreading);
\draw (spreading) -- node[above]{$\underbar x$} (output);
\end{scope}
\end{tikzpicture} 

%% file: decoder.tex
\begin{tikzpicture}[every node/.style={scale=0.8}]
\begin{scope}[node distance=2cm,>=angle 90,semithick]
\node (input){};
\node[block] (SIC)[right of=input] {\textbf{SIC}};

\draw[->] (input) -- node[above]  {$\mathbf{Y}$} (SIC);
\node[block] (SD)[right of=SIC,xshift=1cm]{\textbf{ED}};
\node[block] (mmserec)[right of=SD,xshift=1cm]{\textbf{MMSE}}; 
\draw[->] (SIC) --  node[above]{$\mathbf{Y}^t$} (SD);
\draw[->] (SIC) -- ++(0,1) node [above,xshift=3cm]{$\mathbf{Y}^t$}-| (mmserec);
\node[block] (dec1)[right of=mmserec,xshift=1.5cm]{\textbf{Ch. Decoder}}; 
\node[block] (dec2)[above of=dec1,xshift=0cm,yshift=-0.5cm]{\textbf{Ch. Decoder}};
\node (dots)[below of=dec1,yshift=1cm]{$\vdots$};
\node[block] (dec3)[below of=dots,yshift=1cm]{\textbf{Ch. Decoder}};
\node[sblock] (I1)[right of=dec1,xshift=0.5cm]{\textbf{MOD}};
\node[sblock] (I2)[right of=dec2,xshift=0.5cm]{\textbf{MOD}};
\node[sblock] (I3)[right of=dec3,xshift=0.5cm]{\textbf{MOD}};
\node[cnode] (sum)[right of=I1]{+};
\draw[->] (sum) -- ++(0,-3.5) node[above,xshift=-7cm]{$\widetilde{\mathcal{D}}^t$}-| (SIC);
\draw[->] (mmserec.10) --node[above]{$\underline{\beta}_1^t$} (dec2.180); 
\draw[->] (mmserec.0) -- node[above]{$\underline{\beta}_2^t$}(dec1.180);
\draw[->] (mmserec.350) --node[above,yshift=0.2cm]{$\underline{\beta}_K^t$} (dec3.180);
\draw[->] (I1.0) -- (sum.180); 
\draw[->] (I2.0) -- (sum.160);
\draw[->] (I3.0) -- (sum.200);
\draw[->] (dec1.0) -- node[above]{$\widehat{\underbar v}_2^t$}(I1.180); 
\draw[->] (dec2.0) --node[above]{$\widehat{\underbar v}_1^t$} (I2.180);
\draw[->] (dec3.0) -- node[above]{$\widehat{\underbar v}_{K}^t$}(I3.180);

\draw[->] (SD) -- node[above]{$\mathcal{D}^t$} (mmserec);

\end{scope}
\end{tikzpicture} 

%% file: comparison.tex
\begin{tikzpicture}
\definecolor{mycolor1}{rgb}{0.63529,0.07843,0.18431}%
\definecolor{mycolor2}{rgb}{0.00000,0.44706,0.74118}%
\definecolor{mycolor3}{rgb}{0.00000,0.49804,0.00000}%
\definecolor{mycolor4}{rgb}{0.87059,0.49020,0.00000}%
\definecolor{mycolor5}{rgb}{0.00000,0.44700,0.74100}%
\definecolor{mycolor6}{rgb}{0.74902,0.00000,0.74902}%

\begin{axis}[%
font=\small,
width=7cm,
height=5cm,
scale only axis,
every outer x axis line/.append style={white!15!black},
every x tick label/.append style={font=\color{white!15!black}},
xmin=25,
xmax=175,
xtick = {25,50,100,...,175},
xlabel={Number of active users $\Ka$},
xmajorgrids,
every outer y axis line/.append style={white!15!black},
every y tick label/.append style={font=\color{white!15!black}},
ymin=0.5,
ymax=2,
ytick = {0.5,1,...,2},
ylabel={Required $E_b/N_0$ (dB)},
ymajorgrids,
legend style={at={(0,1)},anchor=north west, draw=black,fill=white,legend cell align=left}
]

\addplot [color=mycolor5,solid,line width=2.0pt,mark size=1.4pt,mark=square,mark options={solid}]
  table[row sep=crcr]{
 25	0.55\\
50	0.6\\
75	0.7\\
100	0.75\\
125 1.15\\
150 1.9\\
};
\addlegendentry{$n_{\mathrm{s}}=29,n_{\mathrm{c}}=1024$};

\addplot [color=mycolor6,solid,line width=2.0pt,mark size=1.4pt,mark=square,mark options={solid}]
  table[row sep=crcr]{25	0.7\\
50	0.75\\
75	0.75\\
100	0.9\\
125	1.25\\
150	1.5\\
175	1.95\\
};
\addlegendentry{$n_{\mathrm{s}}=59,n_{\mathrm{c}}=512$};
\end{axis}
\end{tikzpicture}

%% file: results_ICC.tex
\begin{tikzpicture}
\definecolor{mycolor1}{rgb}{0.63529,0.07843,0.18431}%
\definecolor{mycolor2}{rgb}{0.00000,0.44706,0.74118}%
\definecolor{mycolor3}{rgb}{0.00000,0.49804,0.00000}%
\definecolor{mycolor4}{rgb}{0.87059,0.49020,0.00000}%
\definecolor{mycolor5}{rgb}{0.00000,0.44700,0.74100}%
\definecolor{mycolor6}{rgb}{0.74902,0.00000,0.74902}%

\begin{axis}[%
font=\small,
width=7cm,
height=7cm,
scale only axis,
every outer x axis line/.append style={white!15!black},
every x tick label/.append style={font=\color{white!15!black}},
xmin=25,
xmax=300,
xtick = {25,50,100,...,300},
xlabel={Number of active users $\Ka$},
xmajorgrids,
every outer y axis line/.append style={white!15!black},
every y tick label/.append style={font=\color{white!15!black}},
ymin=0,
ymax=10,
ytick = {0,2,...,10},
ylabel={Required $E_b/N_0$ (dB)},
ymajorgrids,
legend style={at={(0,1)},anchor=north west, draw=black,fill=white,legend cell align=left}
]

\addplot [color=black,dotted,line width=2.0pt]
  table[row sep=crcr]{
 25	0.25\\
50	0.3\\
75	0.35\\
100	0.4\\
125	0.45\\
150	0.5\\
175	0.55\\
200	0.6\\
225	0.95\\
250	1.25\\
275	1.55\\
300	1.8\\
};
\addlegendentry{Random Coding \cite{polyanskiy2017perspective}};

\addplot [color=mycolor2,solid,line width=2.0pt,mark size=1.4pt,mark=square,mark options={solid}]
  table[row sep=crcr]{25	3.18\\
50	3.52\\
75	4.64\\
100	5.61\\
125	5.85\\
150	6.46\\
175	6.72\\
200	7.41\\
225	7.6772\\
250	8.3217\\
275	8.8428\\
300	9.352\\
};
\addlegendentry{SIC T=4 \cite{vem2019user}};

\addplot [color=mycolor3,solid,line width=2.0pt,mark size=1.4pt,mark=o,mark options={solid}]
  table[row sep=crcr]{
  25  2\\
50	2.1\\
75	2.2\\
100	2.41\\
125	2.57\\
150	2.81\\
175	3\\
200 3.4\\
225 3.88\\
250 4.36\\
275 4.87\\
300 5.35\\
};
\addlegendentry{Sparse IDMA \cite{pradhan2019joint}};

\addplot [color=mycolor4,solid,line width=2.0pt,mark size=1.4pt,mark=triangle,mark options={solid}]
  table[row sep=crcr]{
  25  0.73\\
50	1.15\\
75	1.63\\
100	2.06\\
125	2.48\\
150	2.99\\
175	3.43\\
200 3.89\\
225 4.41\\
250 4.91\\
275 5.26\\
300 5.49\\
};
\addlegendentry{IRSA + Polar Code \cite{marshakov2019polar}}
\addplot [color=mycolor1,solid,mark=*,line width=2.0pt]
  table[row sep=crcr]{
  2     0.3\\
  10    0.5\\
 25	0.55\\
50	0.6\\
75 0.7\\
100 0.75\\
125 1.15\\
150 1.5\\
175 2\\
200 2.7\\
225 3.5\\
250 4.3\\
};
\addlegendentry{Proposed Scheme};
\end{axis}

\end{tikzpicture}%